\documentclass[12pt]{article}
\usepackage{graphicx}
\usepackage{multicol} 
\usepackage{color}
\usepackage{amsmath}
\usepackage{amssymb}
\usepackage{amsfonts}

\definecolor{rosso}{cmyk}{0,1,1,0.4}
\definecolor{rossos}{cmyk}{0,1,1,0.55}
\definecolor{rossoc}{cmyk}{0,0.5,1,0.2}
\definecolor{blu}{cmyk}{1,1,0,0.3}
\definecolor{blus}{cmyk}{1,1,0,0.6}
\definecolor{blucc}{cmyk}{1,0.4,0.2,0}
\definecolor{viola}{cmyk}{0,1,0,0.6}
\definecolor{viola2}{cmyk}{0,1,0.2,0.6}
\definecolor{verde}{cmyk}{0.92,0,0.59,0.25}
\definecolor{verdec}{cmyk}{0.92,0,0.59,0.15}
\definecolor{verdes}{cmyk}{0.92,0,0.59,0.4}
\font\tenrsfs=rsfs10 at 12pt
\font\sevenrsfs=rsfs7
\font\fiversfs=rsfs5
\newfam\rsfsfam
\textfont\rsfsfam=\tenrsfs
\scriptfont\rsfsfam=\sevenrsfs
\scriptscriptfont\rsfsfam=\fiversfs
\def\mathscr#1{{\fam\rsfsfam\relax#1}}

\newcommand{\eq}[1]{~{\rm (\ref{eq:#1})}}

\def\circa#1{\,\raise.3ex\hbox{$#1$\kern-.75em\lower1ex\hbox{$\sim$}}\,}

\newcommand{\beq}{\begin{equation}}
\newcommand{\eeq}{\end{equation}}
\newcommand{\bea}{\begin{eqnarray}}
\newcommand{\eea}{\end{eqnarray}}
\newcommand{\diag}{\hbox{diag}\,}

\def\circa#1{\,\raise.3ex\hbox{$#1$\kern-.75em\lower1ex\hbox{$\sim$}}\,}
\makeatletter

\def\art{\@ifnextchar[{\eart}{\oart}}
\def\eart[#1]#2#3#4#5#6{{\rm #2}, {\em #3 \rm #4} {\rm (#6) #5} ({#1})}
\def\hepart[#1]#2{{\rm #2, #1}}
\newcommand{\oart}[5]{{\rm #1}, {\em #2 \rm #3} {\rm (#5) #4}}

\newcounter{alphaequation}[equation]
\def\thealphaequation{\theequation\hbox to
0.6em{\hfil\alph{alphaequation}\hfil}}
\def\eqnsystem#1{
\def\@eqnnum{{\rm (\thealphaequation)}}
\def\@@eqncr{\let\@tempa\relax \ifcase\@eqcnt \def\@tempa{& & &} \or
  \def\@tempa{& &}\or \def\@tempa{&}\fi\@tempa
  \if@eqnsw\@eqnnum\refstepcounter{alphaequation}\fi
\global\@eqnswtrue\global\@eqcnt=0\cr}
\refstepcounter{equation} \let\@currentlabel\theequation \def\@tempb{#1}
\ifx\@tempb\empty\else\label{#1}\fi
\refstepcounter{alphaequation}
\let\@currentlabel\thealphaequation
\global\@eqnswtrue\global\@eqcnt=0 \tabskip\@centering\let\\=\@eqncr
$$\halign to \displaywidth\bgroup \@eqnsel\hskip\@centering
$\displaystyle\tabskip\z@{##}$&\global\@eqcnt\@ne
\hskip2\arraycolsep\hfil${##}$\hfil& \global\@eqcnt\tw@\hskip2\arraycolsep
$\displaystyle\tabskip\z@{##}$\hfil
\tabskip\@centering&\llap{##}\tabskip\z@\cr}
\def\endeqnsystem{\@@eqncr\egroup$$\global\@ignoretrue} \makeatother

\newlength{\myem}
\newcommand{\sep}[1]{#1}
\newcounter{mysubequation}[equation]
\renewcommand{\themysubequation}{\alph{mysubequation}}
\newcommand{\mytag}{\stepcounter{mysubequation}%
\tag{\theequation\protect\sep{\themysubequation}}}
\newcommand{\globallabel}[1]{\refstepcounter{equation}\label{#1}}

\newcommand{\MeV}{\,\hbox{\rm MeV}}
\newcommand{\eV}{\,\hbox{\rm eV}}

\newcommand{\nub}{\bar\nu}
\newcommand{\nue}{\nu_e}
\newcommand{\nueb}{\bar\nu_e}
\newcommand{\numu}{\nu_\mu}
\newcommand{\numub}{\bar\nu_\mu}
\newcommand{\nutau}{\nu_\tau}
\newcommand{\nutaub}{\bar\nu_\tau}

\begin{document}

\thispagestyle{empty}

\begin{flushright}
{
astro-ph/0608206\\

}
\end{flushright}
\vspace{1cm}

\begin{center}
{\LARGE \bf 
Sterile neutrinos, lepton asymmetries,\\[0.2cm] primordial elements:
how much of each?}\\[1cm]

{
{\large\bf Yi-Zen Chu}$^a$, 
{\large\bf Marco Cirelli}$^a$
}  
\\[7mm]
{\it $^a$ Physics Department - Yale University, New Haven, CT 06520, USA}\\[3mm]
\vspace{1cm}
{\large\bf Abstract}
\end{center}
\begin{quote}
{\large\noindent
We investigate quantitatively the extent to which having a primordial leptonic asymmetry $(n_{\nu} \neq n_{\bar\nu})$ relaxes the bounds on light sterile neutrinos imposed by BBN and LSS. We adopt a few assumptions that allow us to solve the neutrino evolution equations over a broad range of mixing parameters and asymmetries. For the general cases of sterile mixing with the electron or muon neutrino, we identify the regions that can be reopened. For the particular case of a LSND-like sterile neutrino, soon to be rejected or confirmed by MiniBooNE, we find that an asymmetry of the order of $10^{-4}$ is needed to lift the conflicts with cosmology.}
\end{quote}



\setcounter{page}{1}
\setcounter{footnote}{0}

\section{Introduction}

Cosmology currently poses some of the most stringent bounds on the properties of sterile neutrinos $\nu_s$ -- Standard Model (SM) gauge group singlet fermions -- namely, their mass and mixing parameters with their active counterparts $\nu_e$, $\nu_\mu$ and $\nu_\tau$.

In particular, by comparing the predictions of the yield of primordial light elements with observations, Big Bang Nucleosynthesis (BBN) efficiently constrains the possible manifestations of sterile neutrinos. Having a greater than 3 effective number of neutrinos would increase the rate of expansion of the universe and leads to larger $^4$He and Deuterium abundances. Even if steriles are produced only after the decoupling of the active neutrinos from the cosmological plasma, so that the effective number of neutrinos remains roughly constant at 3, they would lead to a depletion of electron neutrinos and anti-neutrinos, which also boosts the amount of $^4$He and Deuterium produced~\cite{historical}.
In addition, observations of the matter power spectrum of the Large Scale Structures (LSS) in the Universe limits the total mass in neutrinos today, including the sterile ones~\cite{LesgourguesPastorReview}.

However, naturally light or cleverly lightened sterile fermions arise in many theories beyond the Standard Model.
On the concrete experimental side, moreover, the result of the LSND experiment~\cite{LSND} still stands. The most common interpretation requires the existence of a sterile neutrino with a mass $\mathcal{O}(1 \eV)$ mixing with both the electron and muon neutrino with angles $\mathcal{O}(10^{-3})$.
Such parameters are solidly ruled out by the constraints of standard BBN and by LSS~\cite{CMSV,cosmonu,di bari,LSS}. 

As well known, the same region of the parameters space is currently being scrutinized by the MiniBooNE experiment~\cite{miniboone}, which is expected to deliver results soon.
Should MiniBooNE confirm the existence of the LSND-like sterile neutrino, this would surely signal the need of a drastic revision of standard cosmology. 
If instead MiniBooNE strengthens the experimental exclusion contour beyond the LSND region, the bound from BBN on this particular mixing configuration is still stronger -- a few orders of magnitude in mass and mixing angle -- and these issues remain: in what ways and how much of the parameter space can be reopened?\footnote{This could also be of importance for the $r-$process nucleosynthesis in SuperNov\ae, as Ref.~\cite{rprocess} finds that the region of successful synthesis in the presence of an active/sterile oscillation has some overlap with the LSND area.}

\medskip

The bounds from cosmology are based on the assumption of ``standard evolution''. They relax considerably if some non standard modification is postulated.
One such modification that has attracted much attention is to invoke the presence of a large primordial asymmetry for one or more of the active neutrinos seas.
Of course, one might expect the value for a primordial asymmetry to be set by that of the baryonic asymmetry $\eta = ( n_{\rm B} - n_{\rm \bar{B}} ) / n_{\gamma} \simeq 6 \cdot 10^{-10}$. To respect charge neutrality, the asymmetry in the charged leptons must be of the same order. But there is no such constraint on the asymmetries of the electrically neutral neutrinos, and they are therefore allowed to take values that can be orders of magnitude larger, up to $\mathcal{O}(10^{-2})$~\cite{bounds on xi}. Indeed, there are many examples of cosmological scenarios that lead to a large primordial neutrino asymmetry while preserving the observed smallness of the baryonic one, such as GUT scenarios, the Affleck-Dine mechanism, Q-balls, etc.~\cite{scenariosLargeL}. From the purely phenomenological point of view, this is thus an important possibility and it has relevant consequences on the sterile neutrino bounds.

The basic idea is the following~\cite{original Foot Volkas}. A non vanishing asymmetry $L_\nu$ produces an additional term in the matter potential for the active neutrinos that is proportional to the asymmetry $L_\nu$ itself. This increases the magnitude of the diagonal entries of the real part of the effective Hamiltonian written in the flavor basis relative to its off diagonal ones, thus suppressing the active neutrino oscillations into the sterile states. Since sterile neutrinos can only be created through active-sterile oscillations, this means they will be less efficiently or not produced at all in the Early Universe.

\medskip

The main goal of this paper is to study quantitatively the extent to which such a mechanism works. 
Some aspects of this issue have been addressed before (see e.g.~\cite{asymm, BargerHiding, dolgov villante, fuller}). We aim at an analysis that includes as much content of the full neutrino evolution equations as possible, such as the established active-active neutrino mixing, and that covers a broad range of parameters. We compute the effects in terms of observable cosmological quantities, such as the yield of the primordial elements, and compare them with the most recent results, which have gone through significant changes.

\medskip

The paper is organized as follows. In Sec.\ref{formalism} we describe the neutrino mixing formalism, identify the parameter spaces we are interested in, and present the evolution equations that we solve. In Sec.\ref{results} we present our results and in Sec.\ref{conclusions} we summarize our conclusions. Appendix \ref{appendix} lays down some relatively standard details that govern the evolution of the temperature of the primordial plasma in the time of interest.

\section{Formalism and computations}
\label{formalism}

In this Section we delineate the neutrino mixing formalism and the computational strategy. The uninterested reader can skip to Sec.\ref{results}: the basic points are that, for any given set of values of the active-sterile mixing parameters ($\Delta m^2_{14}$, $\theta_s$) and primordial leptonic asymmetry $L^0 = (n^0_\nu - n^0_{\bar\nu})/n^0_\gamma$ (taken to be the same for the $\nu_e$, $\nu_\mu$ and $\nu_\tau$ seas, for the sake of definiteness~\footnote{In principle, different primordial asymmetries for the different flavors and even conspiracies that compensate the net effect are possible. Our choice seems more generic and it is made on the basis of a balanced initial condition among flavors at high energies. The equalization discussed in~\cite{bounds on xi} will also eventually occur, albeit it could take place at low temperatures.}), we compute the evolution of the densities of the four species of neutrinos and antineutrinos, from an initial high temperature down to a fraction of an MeV. With these, we follow the evolution of the neutron to proton ratio $n/p$: it determines the yield of $^4$He and Deuterium, which we compare with observations. On top of that, the total amount of produced sterile neutrinos is subjected to the bounds from LSS.

\subsection{Neutrino mixing}
To parameterize the mixing of the sterile neutrino with one specific linear combination of the 3 active states, we adopt the general formalism introduced in Ref.~\cite{CMSV}. The extra (mainly sterile) neutrino has a mass $m_4$. It is allowed to mix, with a mixing angle $\theta_s$, with an arbitrary combination of the flavor eigenstates of the active neutrinos, represented by a complex unit 3-vector $\vec{n}$ 
\begin{equation}
\vec{n}\cdot \vec{\nu} = n_e \nu_e + n_\mu \nu_\mu + n_\tau \nu_\tau.
\end{equation}
The $4\times 4$ neutrino mixing matrix $V$, that relates flavor to mass eigenstates as $\nu_{e,\mu,\tau,s} = V\cdot \nu_{1,2,3,4}$ is expressed in this parameterization by
\begin{equation}
V = \left(
\begin{tabular}{cc}
$1-(1-\cos\theta_{\rm s}) \vec{n}^{\bf{\dagger}} \otimes \vec{n}$ & $\sin\theta_{\rm s} \vec{n}^{\bf{\dagger}}$\\ 
 $-\sin\theta_{\rm s} \vec{n}$ & $\cos\theta_{\rm s}$
 \end{tabular}
 \right)
   \times
   \left( 
   \begin{tabular}{cc}
  $U$ & 0 \\
   0 & 1
\end{tabular}
   \right)
\end{equation}
where $U$ denotes the usual $3\times 3$ mixing matrix: $\nu_{\rm f_\ell} = U_{\ell i} \nu_{\rm m_i}$, with $\ell = e, \mu, \tau$ and $f$ and $m$ denoting flavor and mass eigenstates respectively. It contains the mixing angles $\theta_{12}, \theta_{23}, \theta_{13}$ among the active species. Their values, together with those of the active-active squared  mass differences $\Delta m^2_{12}, |\Delta m^2_{23}|$, are fixed by the solar, atmospheric and terrestrial neutrino experiments (see~\cite{StrumiaVissaniReview} for a recent compilation). For simplicity, we restrict to the case of normal hierarchy ($\Delta m^2_{23} >0$) and assume $\theta_{13} = 0$. Since oscillations are sensitive only to mass differences, without loss of generality we set $m_1 = 0$. This is also the most conservative choice with respect to the bounds from LSS. The oscillation results will be presented in terms of the active-sterile squared mass difference $\Delta m^2_{14}$.

In the following we will focus on the particular cases of sterile neutrino mixing with a pure electron neutrino state (``$e$s mixing'') and with a pure muon neutrino state (``$\mu$s mixing''). These correspond to $\vec{n}_\ell  = (1,0,0)$ and $\vec{n}_\ell  = (0,1,0)$ respectively. We will not address the case of mixing with a pure tau neutrino because it is equivalent to the muon case, as they experience the same matter potential and have the same roles in weak interactions. We will limit our investigation to $\Delta m^2_{14} > 10^{-4}{\rm \ eV}^2 \simeq \Delta m^2_{12}$ for the $e$s mixing case and to $\Delta m^2_{14} > 10^{-2}{\rm \ eV}^2 \simeq \Delta m^2_{23}$  for the $\mu$s mixing case so that the (mainly) sterile state is always heavier than the flavor with which it mixes (often called, in absence of asymmetries, the ``non-resonant case''). 
Finally, we will consider the case of the LSND sterile neutrino. This case requires a mixing with both $\nu_e$ and $\nu_\mu$, and therefore interpolates between the previous two cases we have discussed.

\subsection{Neutrino evolution and BBN computations}

The machinery of Big Bang Nucleosynthesis predicts the abundances of the primordial elements produced by the synthesis that occurs at $T_{\rm BBN}\sim \mathcal{O}(1\ {\rm to}\  0.1)\MeV$. We will focus on $^4$Helium and Deuterium. The resulting abundances are dependent on the baryon asymmetry $\eta = (n_{\rm B}-n_{\rm \bar{B}})/n_{\gamma}$ and on the density of relativistic species in the early Universe -- photons, $e^\pm$ and neutrinos.
Fixing the value of $\eta$ with CMB data\footnote{We are assuming that the neutrino asymmetries that we are introducing have a too small effect on CMB to be able to modify this value, as can be inferred from the analysis in~\cite{xi in CMB}.} to be $\eta = (6.11 \pm 0.27) \cdot 10^{-10}$ (we adopt the value from~\cite{cosmonu}, fully compatible with that from~\cite{WMAP3}) and taking the energy spectrum of photons and $e^\pm$ to be those of the standard Bose-Einstein and Fermi-Dirac ones throughout the entire period of evolution, we need to compute the evolution of the densities of the four species of neutrinos and antineutrinos during the whole period of BBN, from $T_{\rm in} \gg T_{BBN}$ to $T_{\rm fin} \lesssim T_{\rm BBN}$.
This evolution is shaped by several processes occurring to the neutrinos: the active species oscillate among themselves as dictated by the active-active mixing angles and mass differences and by the matter potentials of the primordial plasma; the sterile species are populated by oscillations from the active species as dictated by $\theta_s$ and $\Delta m^2_{14}$; eventually active neutrinos decouple from the thermal bath, just before $e^+e^-$ annihilate into photons. In the rest of this Section we delineate the computational formalism that allows us to follow this evolution.

\subsubsection{Treatment of the neutrino evolution equations}
The most convenient tool to follow the system of neutrinos is that of the density matrix~\cite{OscEU}. Intuitively, its diagonal entries correspond to the populations of different species of neutrinos, while the off-diagonal entries express the quantum superposition of flavors. 
Written in the flavor basis, the evolution equations for the full $4 \times 4$ neutrino matrix $\varrho(p,t)$ and antineutrino matrix $\bar\varrho(p,t)$ with momentum $p$ take into account (i) the vacuum oscillations (active-active and active-sterile), (ii) the matter effects in the primordial plasma, including those due to the lepton asymmetry that we are introducing, (iii) the $\nu e \leftrightarrow \nu e$ scattering reactions and the $\nu \bar{\nu} \leftrightarrow e^+e^-$ annihilation reactions. 
They read~\cite{historical, sigl raffelt, dolgov review}
\globallabel{eq:kineq p nu}
\begin{align}
\dot\varrho = \left( \frac{\partial }{\partial t} -  H p \frac{\partial}{\partial p} \right)\varrho 
 = -i \left[ \mathscr{H},\varrho \right] - \left(\begin{tabular}{c}  scattering \& \\  annihilation,~\cite{sigl raffelt} \end{tabular} \right), \mytag \\
\dot{\bar{\varrho}} = \left( \frac{\partial }{\partial t} -  H p \frac{\partial}{\partial p} \right)\bar\varrho 
 = -i \left[ \mathscr{H},\bar\varrho \right] - \left(\begin{tabular}{c}  scattering \& \\ annihilation,~\cite{sigl raffelt} \end{tabular} \right). \mytag 
\end{align}
The Hubble parameter 
\beq
H = \sqrt{\frac{8 \pi^3 G_N}{3} \wp_{\rm tot}}
\eeq
depends on all the energy densities $\wp$, including those of active and sterile neutrinos and antineutrinos, which are functions of the diagonal entries of the density matrices: $\wp_{\rm tot} = \wp_\gamma + \wp_{e^\pm} + \sum_{i=1}^4 \wp_{\nu_i}( \rho_{ii} ) + \sum_{i=1}^4 \wp_{\nub_i}( \bar{\rho}_{ii} )$. 
The Hamiltonian in matter
\beq 
\mathscr{H} = \mathscr{H}_0 + \diag(\mathscr{V}_e, \mathscr{V}_\mu, \mathscr{V}_\tau, 0)
\eeq	
is composed by the vacuum Hamiltonian in the flavor basis
\beq
\mathscr{H}_0 = \frac{1}{2p} V M^2 V^\dagger \qquad  M^2 = \diag(m_1^2,m_2^2, m_3^2,m_4^2)
\eeq
and the matter potentials $\mathscr{V}_l$ for each flavor~\cite{notzold}
\globallabel{eq:p potentials}
\begin{align}
\begin{split}
\mathscr{V}_e  & = \pm \sqrt{2} G_F n_\gamma \left[ \frac{1}{2} \eta + 2 L_{\nue} + L_{\numu} + L_{\nutau} \right] 
   - \frac{8 \sqrt{2} G_F}{3 M_Z^2} p \big( \langle E_\nu \rangle n_{\nue} + \langle E_{\nub} \rangle n_{\nueb}  \big) \\
   & \quad - \frac{8 \sqrt{2} G_F}{3 M_W^2} p \big( \langle E_e \rangle n_{e} + \langle E_{\bar{e}} \rangle n_{\bar{e}}  \big)
\end{split} \mytag \\
\mathscr{V}_\mu & =  \pm \sqrt{2} G_F n_\gamma \left[ \frac{1}{2} \eta + L_{\nue} + 2 L_{\numu} + L_{\nutau} \right] 
   - \frac{8 \sqrt{2} G_F}{3 M_Z^2} p \big( \langle E_\nu \rangle n_{\numu} + \langle E_{\nub} \rangle n_{\numub}  \big) \mytag \\
\mathscr{V}_\tau & = \pm \sqrt{2} G_F n_\gamma \left[ \frac{1}{2} \eta + L_{\nue} + L_{\numu} + 2 L_{\nutau} \right] 
   - \frac{8 \sqrt{2} G_F}{3 M_Z^2} p \big( \langle E_\nu \rangle n_{\nutau} + \langle E_{\nub} \rangle n_{\nutaub}  \big) \mytag \\
\mathscr{V}_s & = 0 \mytag
\end{align}
where we are making use of the standard definitions of the asymmetry
\beq
\label{def asymmetry}
L_{\nu} = \frac{n_{\nu} - n_{\bar\nu}}{n_\gamma}
\eeq
and of the number density and average energy of a species $x$ with distribution $f_x(p)$ and $g_x$ spin degrees of freedom
\beq
n_x = \int \frac{d^3 p}{(2 \pi)^3} \ g_x \ f_x (p), \qquad \langle E_x \rangle = \frac{1}{n_x} \int \frac{d^3 p}{(2 \pi)^3} \ g_x \ E(p) \ f_x (p).
\eeq
For photons, $E(p) = p$ and $n_\gamma = 2 \zeta(3)/\pi^2\ T^3$. Neutrinos and antineutrinos are distinct and have $g_\nu = g_{\bar\nu} =1$; also, because they are nearly massless relative to the ambient temperature, $E(p) = \sqrt{m_{\nu}^2 + p^2} \approx p$.

The first terms in each expression of the $\mathscr{V}_l$ are the contribution from the asymmetries: the baryon asymmetry $\eta$ and the asymmetries in the background neutrinos of all flavors, with a factor 2 in front of the same-flavor contribution; the plus sign applies to test neutrinos, the minus sign to test antineutrinos. Notice that the leptonic asymmetry will start from its primordial value $L_\nu^0$, one of the free parameters of our analysis, but it is in principle a dynamical quantity, determined by the evolution of the neutrino and antineutrino densities. This might complicate the equations non trivially. We will further comment on this point below. 

The second term is the extra contribution due to the presence in the plasma of background neutrinos of the same flavor as the test neutrino. This term does not vanish even in case of symmetry in the background densities and effectively provides a thermal mass to neutrinos immersed in the bath.  
The third term in the electron neutrino potential similarly comes from the background electrons and positrons of the bath, and it is not there for neutrinos of other flavors. 

The last terms in eq.s\eq{kineq p nu} contain the rates $\Gamma$'s of the $\nu e \leftrightarrow \nu e$, $\nu \nu \leftrightarrow \nu \nu$ scattering reactions and the $\bar\nu \nu \leftrightarrow e^+e^-, \bar\nu \nu$ annihilation reactions.
Their complete form can be found in~\cite{sigl raffelt}. We will provide the explicit form that they take in our formalism below.

\bigskip

The momentum dependent equations eq.\eq{kineq p nu} do not allow for an efficient way of studying the evolution of neutrinos and antineutrinos over a broad range of parameters, which is the goal of our paper. We work therefore in the limit in which the neutrino distribution can be approximated as Fermi-Dirac functions with a negligibly small chemical potential, multiplied by an overall normalization factor that can be interpreted as an average of the available phase space: 
\beq
\label{eq:ansatz}
\varrho(p,t) \to \frac{\rho(t)}{e^{p/T_\nu}+1}, \qquad \bar\varrho(p,t) \to \frac{\bar\rho(t)}{e^{p/T_\nu}+1}
\eeq
Here $\rho$ and $\bar\rho$ are $4 \times 4$ momentum-independent density matrices that evolve with time only.
Even though we are dealing with particle asymmetries, neglecting the chemical potential is a good approximation as long as the asymmetries under consideration are small, as will be in this paper. Quantitatively, one can estimate a priori the typical size of the asymmetries that we will be considering by comparing the contributions of the asymmetry term and the thermal term in eq.\eq{potentials}. For a neutrino of average momentum $p \sim 3.15\ T$, the two become comparable for 
\beq
\label{estimate}
L_\nu \approx 3.15 \frac{7 \pi^4}{180\ \zeta(3)}\left(1+ \frac{1}{c_W^2}\right) \frac{T^2}{M_Z^2} \approx 3\ 10^{-7} \left(\frac{T}{10 \MeV} \right)^2 
\eeq 
In other words, an asymmetry $L_\nu$ of the order of $\mathcal{O}(10^{-6}$) is already enough to give a dominant  contribution to the matter potential (and therefore an important effect on active-sterile oscillations) in the range of temperatures $T\sim {\rm few\  tens}\, \MeV$ and range of masses $\Delta m^2_s < {\rm few}\ \eV^2$ in which we are interested, as one can verify~\cite{dolgov villante} that the production of sterile neutrinos (in absence of asymmetries) dominantly occurs at a temperature 
\beq
\label{eq:Tprod}
T_{{\rm prod}\ \nu_s}\approx 10 \MeV \left(\frac{3T_\nu}{E_\nu} \right)^{1/3} \left( \cos2\theta_s \right)^{1/6} \left(\frac{\Delta m^2_s}{\eV^2} \right)^{1/6}
\eeq
Since an asymmetry $L_\nu$ translates into a dimensionless chemical potential $\xi = \mu/T$ via
\beq
L_\nu = \frac{\pi^2}{12\zeta(3)} \left(\frac{T_\nu}{T} \right)^3 \left( \xi+\frac{\xi^3}{\pi^2} \right),
\eeq
for small values such as $10^{-6}$, we have $\xi \sim L_\nu$, which requires a negligible modification to the pure FD distribution. 
For much larger asymmetries our assumption would be questionable. 

Other kinds of deviations from a pure FD distribution could occur during the evolution, possibly due to the conversions into sterile neutrinos themselves. These have been shown e.g. to be not negligible for $\Delta m^2_{\rm s} \lesssim 10^{-8}\ {\rm eV}^2$~\cite{kirilova}, which is however a region far from that of our investigation. More recently, Ref.s~\cite{fuller} have shown that matter resonances post weak decoupling might leave both active and sterile neutrinos with a highly non-thermal spectrum for some choices of neutrino parameters and energies. This appears to be the case in particular for values of the initial asymmetries larger than those that we will investigate. However, a fully momentum-dependent evolution analysis, which is beyond the scope of this paper, would be necessary to assess the importance of this effect in the entire range of asymmetries and mixing parameters that we consider.

\bigskip

With the assumptions outlined above, the evolution equations for the neutrino and antineutrino $\varrho$ matrices can be integrated over the momentum $p$ and reduced to the corresponding equations for the $\rho$ matrices. Explicitly, they read

\globallabel{eq:kineq nu}
\begin{align}
\dot\rho=\frac{dT}{dt}\frac{d\rho}{dT}=-i \left[ \mathcal{H},\rho \right] 
- \frac{1}{2} \Gamma^0 \bigg(\left\{G_{\rm s}^2 , \rho-\rho^{\rm eq} \right\} - 2 G_{\rm s} (\rho-\rho^{\rm eq}) G_{\rm s} \phantom{xxxx} \nonumber \\
 + \left\{G_{\rm a}^2 , (\rho-\rho^{\rm eq}) \right\} + 2 G_{\rm a} (\bar\rho-\bar\rho^{\rm eq}) G_{\rm a}\bigg),
\mytag \\
\dot{\bar\rho}=\frac{dT}{dt}\frac{d\rho}{dT}=-i \left[ \mathcal{H},\bar\rho \right] 
- \frac{1}{2} \Gamma^0 \bigg( \left\{ G_{\rm s}^2 , \bar\rho-\bar\rho^{\rm eq} \right\} - 2 G_{\rm s} (\bar\rho-\bar\rho^{\rm eq}) G_{\rm s} \phantom{xxxx} \nonumber \\
+ \left\{ G_{\rm a}^2 , (\bar\rho-\bar\rho^{\rm eq}) \right\} + 2 G_{\rm a} (\rho-\rho^{\rm eq}) G_{\rm a} \bigg),
\mytag
\end{align}
where now
\beq
\mathcal{H} = \mathcal{H}_0+\diag(\mathcal{V}_e, \mathcal{V}_\mu, \mathcal{V}_\tau, 0),
\eeq
\beq
\mathcal{H}_0 = \frac{1}{2}\frac{\pi^2}{18\ \zeta(3) T_\nu} V M^2 V^\dagger,
\eeq
\globallabel{eq:potentials}
\begin{align}
\begin{split}
\mathcal{V}_e & = \pm \sqrt{2} G_F n_\gamma \left[ \frac{1}{2} \eta  +\frac{3}{8} \left(\frac{T_\nu}{T} \right)^3 \Big( 2 (\rho_{ee} -\bar\rho_{ee}) + (\rho_{\mu\mu} -\bar\rho_{\mu\mu})+ (\rho_{\tau\tau} -\bar\rho_{\tau\tau}) \Big) \right] \\
 & \quad -\frac{196\ \pi^2}{180} \frac{\zeta(4)}{\zeta(3)}\sqrt{2}G_F \frac{T_\nu}{M_W^2}\left[ T^4 +\frac{1}{4}T_\nu^4 \cos^2\theta_w (\rho_{ee} +\bar\rho_{ee}) \right], 
\end{split} \mytag \\
\begin{split}
\mathcal{V}_\mu & = \pm \sqrt{2} G_F n_\gamma \left[ \frac{1}{2} \eta  +\frac{3}{8} \left(\frac{T_\nu}{T} \right)^3 \Big( (\rho_{ee} -\bar\rho_{ee}) + 2 (\rho_{\mu\mu} -\bar\rho_{\mu\mu})+ (\rho_{\tau\tau} -\bar\rho_{\tau\tau}) \Big) \right]  \\
 & \quad -\frac{196\ \pi^2}{180} \frac{\zeta(4)}{\zeta(3)}\sqrt{2}G_F \frac{T_\nu}{M_W^2}\frac{1}{4}T_\nu^4 \cos^2\theta_w (\rho_{\mu\mu} +\bar\rho_{\mu\mu}), 
\end{split} \mytag \\
\begin{split}
\mathcal{V}_\tau & = \pm \sqrt{2} G_F n_\gamma \left[ \frac{1}{2} \eta  +\frac{3}{8} \left(\frac{T_\nu}{T} \right)^3 \Big( (\rho_{ee} -\bar\rho_{ee}) + (\rho_{\mu\mu} -\bar\rho_{\mu\mu})+ 2 (\rho_{\tau\tau} -\bar\rho_{\tau\tau}) \Big) \right]  \\
 & \quad -\frac{196\ \pi^2}{180} \frac{\zeta(4)}{\zeta(3)}\sqrt{2}G_F \frac{T_\nu}{M_W^2}\frac{1}{4}T_\nu^4 \cos^2\theta_w (\rho_{\tau\tau} +\bar\rho_{\tau\tau}). 
\end{split} \mytag
\end{align}
$\sin\theta_{w} = 0.2312$ is the weak mixing angle.

In these equations we have denoted
\beq
\Gamma^0 = G_F^2 T_\nu^5
\eeq
while $G_{\rm s}$ and $G_{\rm a}$ are diagonal matrices that contain the numerical coefficients for the scattering and annihilation processes for the different flavors. Explicitly
\beq
G_{\rm s, a} = {\rm diag} \big(\gamma_{\rm s, a}^e, \gamma_{\rm s, a}^\mu, \gamma_{\rm s, a}^\tau, 0\big) \qquad {\rm with}\ \ \ \begin{array}{ll} 
(\gamma_{\rm s}^e)^2 = 3.06, & (\gamma_{\rm a}^e)^2 = 0.50, \\ 
(\gamma_{\rm s}^{\mu,\tau})^2 = 2.22, & (\gamma_{\rm a}^{\mu,\tau})^2 = 0.28 . \end{array} 
\eeq
The numerical coefficients $\gamma$ can be determined from the expressions for the reaction rates reported e.g. in~\cite{enqvist}.

Our choice for the scattering and annihilation terms deserves some comment. They are based on eq.(\ref{eq:kineq p nu}) integrated over all momenta. In particular, its form can be obtained by considering the density matrix as a function of its difference from its equilibrium value, $\rho = \rho^{\rm eq} + \delta \rho$ and $\bar\rho = \bar\rho^{\rm eq} + \delta \bar\rho$, where $\rho^{\rm eq} = {\rm diag} \big(1 + 4/3 L_{\nu_e}, 1 + 4/3 L_{\nu_\mu}, 1 + 4/3 L_{\nu_\tau}, 1 + 4/3 L_{\nu_s} \big)$, $\bar\rho^{\rm eq} = {\rm diag} \big(1 - 4/3 L_{\nu_e}, 1 - 4/3 L_{\nu_\mu}, 1 - 4/3 L_{\nu_\tau}, 1 - 4/3 L_{\nu_s} \big)$. Strictly speaking, the equilibrium density matrices are the solutions that would make the RHS of eq.(\ref{eq:kineq p nu}) vanish, which means $\rho^{\rm eq}$ and $\bar\rho^{\rm eq}$ must be proportional to the identity matrix: $L_{\nu_e} = L_{\nu_\mu} = L_{\nu_\tau} = L_{\nu_s} \equiv L_{\nu}$. However, at high temperatures when oscillations are heavily suppressed $\mathcal{H}$ becomes nearly diagonal, and equilibrium can be achieved with simply a diagonal matrix with possibly different asymmetries $L_{\nu_i}$. Furthermore, because the only occurences of $\rho^{\rm eq}$ and $\bar\rho^{\rm eq}$ are those inside the anti-commutator of the annihilation terms\footnote{Because the matrix $G_{\rm s}$ is diagonal, the scattering terms do not depend on $\rho^{\rm eq}$ and $\bar\rho^{\rm eq}$, as can verified by writing them as nested commutators $(1/2) \Gamma^0 \left( - \{ G_s^2, \rho - \rho^{\rm eq} \} + 2 G_s (  \rho - \rho^{\rm eq} ) G_s \right) = (1/2) \Gamma^0 [G_s,[\rho - \rho^{\rm eq}, G_s]] = (1/2) \Gamma^0 [G_s,[\rho, G_s]]$.}, and because $(G_{\rm s, a})_{ss} = 0$ the elements $(\rho^{\rm eq})_{ss}$ and $(\bar\rho^{\rm eq})_{ss}$ do not appear anywhere in the equations and hence can be set to zero. This is consistent with the fact that sterile neutrinos can be driven to equilibrium with other neutrino flavors only through oscillations.

In general, the effect of the scattering and annihilation terms is to drive the density matrices to their equilibrium form, effectively replenishing the active neutrino populations at the expenses of the $e^+e^-$ bath and damping the flavor superpositions.
When the rates $\Gamma_{\nu_{f_i}}$ (that strongly depend on $T$) are overwhelmed by the expansion of the Universe ($\Gamma_{\nu_{f_i}} < H$) the so called neutrino freeze-out occurs, though neutrinos still take part in the $n \leftrightarrow p$ reactions (see below). 
It is not difficult to see that the diagonal entries in the scattering terms vanish, hence the fact that scatterings do not contribute to the change in the number density of neutrinos and antineutrinos is built in the equations. On the contrary, the off-diagonal entries are affected by the total scattering rate because both elastic and inelastic scattering play a role in damping out the coherence between different neutrino flavors.

Moreover, we note that the equations by themselves guarantee the conservation of the total lepton number, given by the algebraic sum of the lepton asymmetries of the 3 active and 1 sterile flavors (which can individually vary). To see this, subtract the $\bar\nu$ from the $\nu$ equations to get 
\beq
\dot\rho-\dot{\bar\rho} = -i[\mathcal{H}, \rho ] + i[\mathcal{H}, \bar{\rho} ] -\frac{1}{2} \Gamma^0 \left( \begin{tabular}{l} $[(\rho-\rho^{\rm eq})G_{\rm s} - {\rm h.c.},G_{\rm s}] + \bar\nu \ \ {\rm terms}$ +\\ 
$[(\rho-\rho^{\rm eq})G_{\rm a} - {\rm h.c.},G_{\rm a}] + \bar\nu \ \ {\rm terms}$ \end{tabular} \right).
\eeq
The commutator structure ensures that the total lepton number is strictly conserved: 
\beq
\frac{d}{dt}\sum_{i=1}^4(\rho_{ii}-\bar\rho_{ii}) = {\rm Trace}([\ldots,\ldots])=0 .
\eeq

We finally stress that the popular (and simpler) anticommutator approximation of the scattering and annihilation terms (see e.g. eq.(283) of~\cite{dolgov review}) cannot be applied to our investigation of lepton asymmetries without additional constraint equations because it does not conserve total lepton number.\footnote{When $L_\nu = 0$, the difference between the anti-commutator approximation and our equations here can most likely be attributed to our inclusion of Pauli blocking factors.}

One last comment is in order. Although the evolution equations presented in eq.(\ref{eq:kineq nu}) treat the asymmetries $\rho_{\alpha\alpha}-\bar\rho_{\alpha\alpha}$ as fully dynamical quantities, for numerical reasons we find it necessary to restrict the analysis to the case where the asymmetry is constant\footnote{As a technical aside: we implemented this by replacing $\bar\rho$ with $\rho - (8/3) L$ in the commutator $-i[\mathcal{H}, \bar\rho]$, where $L = {\rm diag}(L_\nu, L_\nu, L_\nu, 0)$. By subtracting the $\rho$ equations and these modified $\bar\rho$ ones, it is straightforward to show that the diagonal terms are zero.}.
This is a reasonable approximation as long as the evolution itself does not induce drastic modifications. Indeed the latter have been studied in some detail and found to be important in the region of small $\Delta m^2_{\rm s}$ and mixing angle~\cite{dolgov villante, rising of asy}, but not in the main region of our interest. With our simplification, the evolution of neutrinos and antineutrinos are however not fully independent. In the event that a positive signal for a sterile neutrino is found at the MiniBooNE experiment or elsewhere, it might be worthwhile to abandon the above simplification and tackle the numerical problem of solving the evolution equations in their full complexity for the parameters possibly selected by that discovery.


\bigskip

The determination of $\frac{dT}{dt}$ is in principle quite involved and we leave its discussion for Appendix~\ref{appendix}. For the purpose of the above equations, we make use of the standard expression $\dot{T}=-H(\wp_{\nu_{\rm tot}})\ T$, where $H$ contains the temperature dependent total energy density, including that of all neutrinos. The standard definition of $\wp_{\rm tot}$ is recalled in eq.\eq{totalenergydensity}. In the density matrix formalism
\beq
N_\nu = \sum_{\alpha=e,\mu,\tau,s} \frac{1}{2} \left( \rho_{\alpha\alpha}+ \bar\rho_{\alpha\alpha} \right)
\eeq

\bigskip

We therefore solve these evolution equations from an initial temperature of $T = 200 \MeV$, much higher than the typical production temperature of sterile neutrinos for any $\Delta m^2$ in which we are interested (see eq.\eq{Tprod}) to a final temperature $T \ll 1\MeV$. At $200 \MeV$, the abundance of the sterile neutrinos is assumed to be zero, for even if they are produced by some New Physics mechanism at very high scales, they would have been diluted by the reheating of the Universe due to the annihilation of the SM degrees of freedom).\footnote{For the opposite possibility, see~\cite{kirilova2005}.} The abundance of active neutrinos and antineutrinos at the initial temperature are $1 + (4/3)L_\nu$ and $1 - (4/3)L_\nu$ respectively.

\subsubsection{Neutron/proton evolution equations and primordial elements}

After solving the neutrino densities evolution as a function of temperature, we can study the relative neutron/proton abundance, which is the all-important quantity for primordial nucleosynthesis, since they are the building blocks of the nuclei and essentially all neutrons are incorporated into some light element in the process. The neutron abundance at the moment that the synthesis begins practically fixes the proportions of all the products. $n/p$ evolves according to
\begin{equation}
\label{eq:noverp}
\dot{r} \equiv \frac{dT}{dt} \frac{dr}{dT} = \Gamma_{p \to n} (1-r)- r \Gamma_{n \to p}, \qquad
r = \frac{n_n}{n_n+n_p}.
\end{equation}
$\Gamma_{p \to n}$ is the total rate for all the $p \to n$ reactions and $\Gamma_{n \to p}$ for the inverse processes. They depend on the electron neutrino and antineutrino densities computed above. 
Explicitly they are given by (see e.g.~\cite{dolgov review,esposito})

\begin{align}
\label{eq:pnreactions}
\begin{split}
\Gamma_{p \to n}  = & \Gamma_{p \nueb \to n e^+} +\Gamma_{pe^- \to n\nue}+\Gamma_{pe^-\nueb \to n} = \\
= & (1+3g_A^2)G_F^2/ (2\pi^3) \left( \int_{m_e}^\infty {\rm d}E_e\ p^2_\nu E_e p_e f_{\nueb} (p_\nu)\left[ 1-f_{e^+}(E_e) \right] \rfloor_{E_\nu = E_e+\Delta m} \right. \phantom{xxxi}\\
& + \int_0^\infty {\rm d}p_\nu\ p_\nu^2 p_e f_e(E_e)\left[1-f_{\nue}(p_\nu) \right]\rfloor_{E_e=p_\nu+\Delta m} \\
& \left.+ \int_{m_e}^{\Delta m} {\rm d}E_e\ p_\nu^2 E_e p_e f_{\nueb}(p_\nu)f_e(E_e)\rfloor_{p_\nu = \Delta m-E_e} \right),
\end{split}
\end{align}

\begin{align}
\label{eq:npreactions}
\begin{split}
\Gamma_{n \to p} = & \Gamma_{n \to p e^{-} \nueb} +\Gamma_{n \nue \to p e^-} + \Gamma_{n e^+ \to p \nueb} = \\
= & (1+3g_A^2)G_F^2/ (2\pi^3) \left( \int_{m_e}^{\Delta m} {\rm d}E_e\ p_\nu^2 E_e p_e \left[1-f_{\nueb}(p_\nu) \right] \left[1-f_e (E_e) \right] \rfloor_{p_\nu = \Delta m - E_e} \right. \\
 &+ \int_{0}^{\infty} {\rm d}p_\nu\ p_\nu^2 E_e p_e f_{\nue}(p_\nu) \left[1-f_e (E_e) \right] \rfloor_{E_e = p_\nu + \Delta m} \\
 & \left. + \int_{m_e}^{\infty} {\rm d}E_e\ p_\nu^2 E_e p_e f_{e^+}(E_e) \left[1-f_{\nueb} (p_\nu) \right] \rfloor_{p_\nu = E_e + \Delta m} \right),
\end{split}
\end{align}
that we evaluate taking standard FD distributions with vanishing chemical potential for electrons and positrons, and the expressions in eq.\eq{ansatz} for the neutrino distributions. 
Here $\Delta m = m_n - m_p$, the mass difference between neutron and proton, $g_A = - 1.267$ is the axial coupling for the neutron and the overall coefficient is best determined in terms of the standard neutron lifetime in absence of a thermal bath 

\beq
\tau_n^{-1} = \frac{(1+3g_A^2)G_F^2}{2\pi^3} m_e^5 \lambda_0, \qquad \lambda_0 = \int_1^{\Delta m/m_e} {\rm d}\epsilon\ \epsilon \sqrt{\epsilon^2-1}\left(\epsilon-\frac{\Delta m}{m_e}\right)^2 \simeq 1.633,
\eeq
experimentally measured to be (885.7 $\pm$ 0.8) s~\cite{PDG}. 

\bigskip

The determination of $\frac{dT}{dt}$ is affected by the several phenomena that go on in the range $T\sim 1 \MeV$ which is under examination. In particular, on top of the expansion that cools the Universe, there is the production of possible extra degrees of freedom (the sterile neutrinos) and the heating of the thermal bath due to $e^+ e^-$ annihilations. We deal with this in Appendix~\ref{appendix}.

\bigskip

From the final value of $n/p$ a set of nuclear physics processes determine the predictions for the synthesis of the light elements. To perform this step we use the procedure briefly described in~\cite{CMSV}, calibrated to agree with more refined codes.
The predictions are finally tested against the observations to draw constraints on the allowed modifications. We focus on the $^4$He and on the Deuterium abundance: the observational determinations of both quantities are plagued by controversial systematic uncertainties that, especially in the case of $^4$He, seem to restrict the margin for improvement; the case of Deuterium may have brighter prospects in this regard. 
We adopt the latest PDG recommended values~\cite{PDGSarkar} for the observational determinations of: 
\beq\begin{array}{c}\label{eq:pD}
 Y_p = 0.249\pm 0.009,  \\
 \displaystyle
 Y_{\rm D}/Y_{\rm H} =  (2.78 \pm 0.29)\cdot10^{-5} ,
\end{array}\eeq
where $Y_X  \equiv n_{X}/n_B$ and $Y_p$ is the traditional notation for $Y_{^4{\rm He}}$. These values are dominated by the results in~\cite{OliveSkillman} and~\cite{Kirkman} and are considered quite conservative (see~\cite{proceeding} for a slightly expanded account). For this reason we adopt
\beq\begin{array}{c}\label{eq:maxpD}
 Y_p \le 0.258,  \\
 \displaystyle
 Y_{\rm D}/Y_{\rm H} \le 3.07 \cdot 10^{-5} 
\end{array}\eeq
as current robust observational upper bounds.

\subsection{Large Scale Structure bounds}
\label{sec:LSS}
The measurements of the matter power spectrum of Large Scale Structures from a variety of techniques, combined with CMB data, allow one to constrain the amount of energy density in neutrinos today, including the sterile one (see e.g.~\cite{LesgourguesPastorReview}). The bound translates into a function of the mass and abundance of the sterile neutrinos~\cite{omeganu}. In our case these two quantities are related, because the neutrino mass differences (together with the other parameters) determine the final abundance. If the sterile neutrino is fully populated by oscillations, its mass contributes fully to the computation of the total energy density. If it is not efficiently produced, the bound on its mass can be somewhat lifted. This is what we will study quantitatively below.

In the regimes in which we are interested, the bound can be expressed to a good approximation in terms of the quantity
\beq
\Omega_\nu h^2 = \frac{1/2\, {\rm Tr}[M \cdot (\rho+\bar\rho)]}{93.5\ {\rm eV}}
\eeq
where $M$ is the 4-neutrino mass matrix introduced before and the neutrino densities $\rho, \bar\rho$ are taken at late cosmological time, in our case the end of the evolution. We assume here that its value is not significantly modified at low $T$. Here as usual $h = H_0/(100\ {\rm km}/{\rm s\ Mpc})$. 
We adopt the constraint
\beq
\label{eq:maxOmega}
\Omega_\nu h^2 \le 0.8 \cdot 10^{-2},
\eeq
based on~\cite{cosmonu}, which is in agreement with most of the recent analysis~\cite{summnu}.

\section{Results}
\label{results}

As discussed above, we solve numerically the neutrino and neutron/proton evolution equations in a broad range of values of the active-sterile mixing parameters and of the primordial asymmetry $L_\nu$. We focus on the case of a negative value for the primordial asymmetry, motivated by the following. Because steriles are produced solely through oscillations from active states, their rate of production up to dimensionless factors goes as $\dot{\rho}_{\rm ss} \sim \sin^2(2 \theta_{\rm matter}) \Gamma^0$, where for small angles $\sin^2(2 \theta_{\rm matter}) \approx \tan^2(2 \theta_{\rm matter}) = \sin^22\theta_s/ ( \cos2\theta_s - 2 (E_\nu / \Delta m^2_s) ( \mathcal{V}_{\rm thermal} + \mathcal{V}_{\rm asymm} ) )^2$~\cite{dolgov review,BD}. From eq.(\ref{def asymmetry}) and eq.(\ref{eq:potentials}), a negative asymmetry would make all terms in the denominator of $\tan^2(2 \theta_{\rm matter})$ positive, leading to a smaller effective mixing angle and hence a reduced production rate.
For large values of $L_\nu$, the asymmetry term, which changes relative sign for $\bar\nu$, could become dominant over the thermal term and allow for resonant transitions in the anti-neutrino channel to take place. As part of our approximations discussed above, we are neglecting these effects.

We see noticeable suppression of $\nu_s$ production when the absolute value of the asymmetry is above $\mathcal{O}(10^{-7})$, consistent with the estimates in eq.(\ref{estimate}). The suppression causes the population of sterile neutrinos to rise later in the evolution, or to be kept at zero in the limiting case.

\subsection{Bounds on sterile mixing parameters}

\begin{figure}[t]
$$
\includegraphics[width=0.48\textwidth]{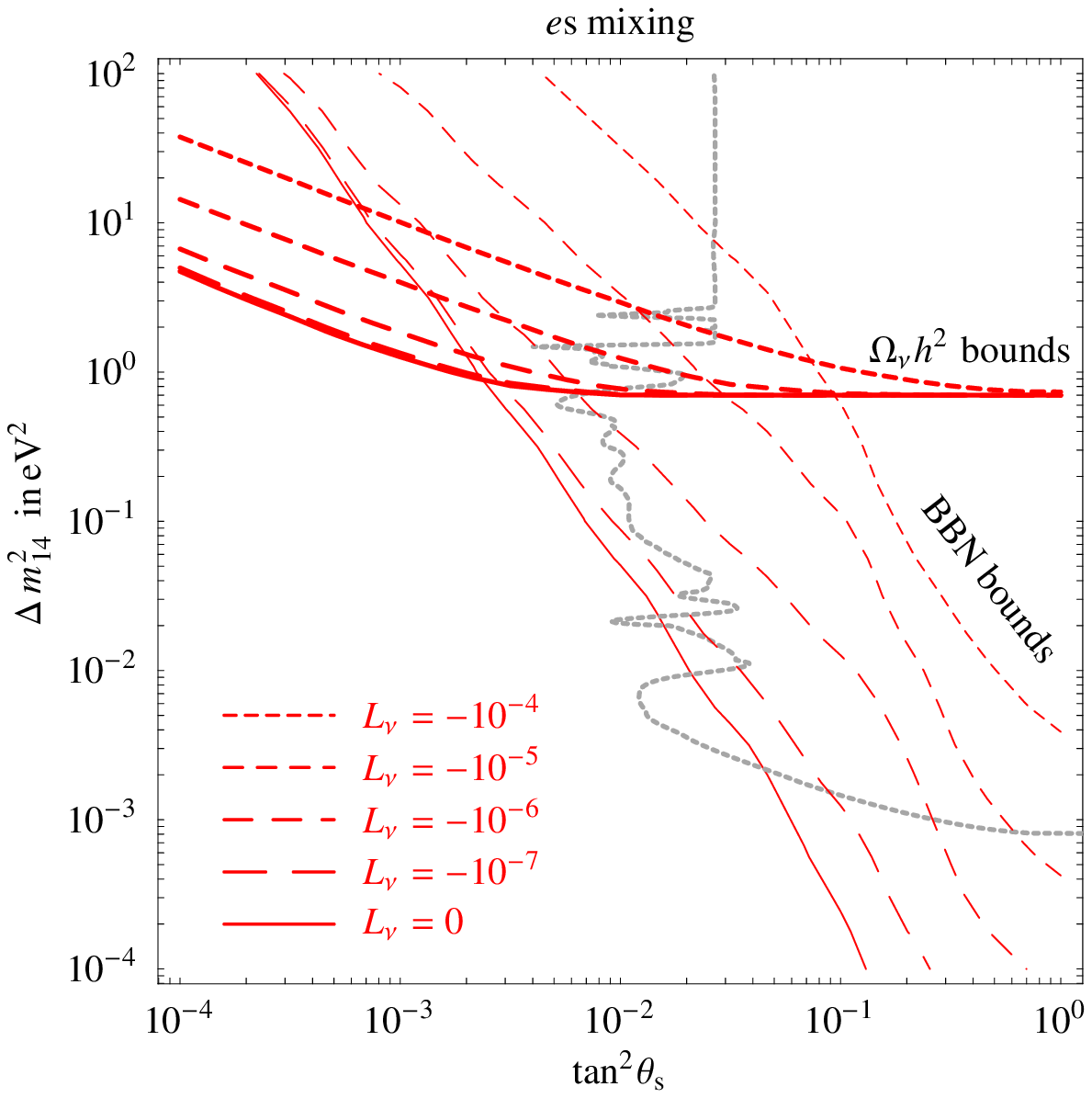}\qquad
\includegraphics[width=0.48\textwidth]{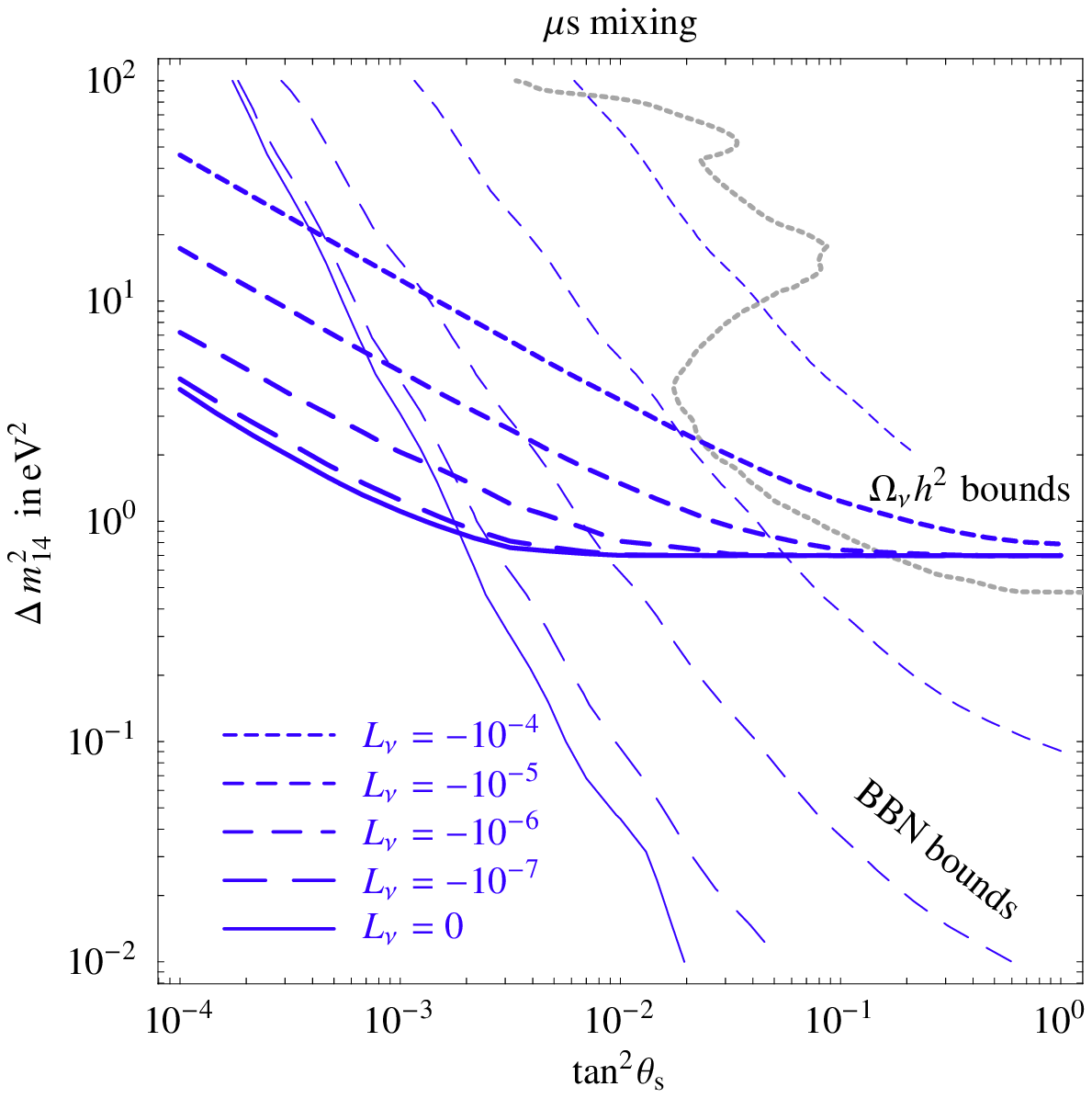}$$
\caption[X]{\label{fig:esmus}\small The bounds from BBN and LSS on sterile neutrinos mixing with electron neutrinos (left panel) and with muon neutrinos (right panel), for different values of the primordial asymmetry $L_\nu$. The regions above and to the right of the thin lines are excluded by BBN because they correspond to $Y_p \ge 0.258$. The regions above and to the right of the thick lines are excluded by LSS because they correspond to $\Omega_\nu \ge 0.8\ 10^{-2}$. The gray dotted lines enclose the excluded regions from a combination of reactor and accelerator experiments (see Sec.\ref{sec:LSND}). Solar and atmospheric experiments also exclude similar regions, reported in~\cite{CMSV}.}
\end{figure}

Fig.s~\ref{fig:esmus} show the bounds on the planes ($\Delta m^2_{14}, \tan^2\theta_{\rm s}$) for the mixing cases $e$s and $\mu$s, defined in Sec.\ref{formalism}. We report the contours for $Y_p = 0.258$ and $\Omega_\nu h^2 = 0.8\ 10^{-2}$, so that the portions above and to the right of these contours are to be considered excluded according to eq.(\ref{eq:maxpD}) and eq.(\ref{eq:maxOmega}). We find that the bound from Deuterium abundance is less constraining than that from $^4$He, so we do not plot it. The determination of primordial Deuterium in eq.(\ref{eq:pD}) has to be improved by at least a factor of 3 in the uncertainty for it to become competitive in this respect.

Focusing first on the BBN bounds, the contours for $L_\nu=0$ have perfect agreement with those in the literature~\cite{CMSV, dolgov villante}, apart from the fact that we used the most recent cosmological values. Introducing the primordial asymmetry $L_\nu$ causes the contours to move upwards and to the right, as expected, thus reopening a portion of the parameter space. The bounds in the $e$s mixing case for large asymmetries reach lower values of $\Delta m^2_{14}$ than those in the $\mu$s cases. This is because the depletion of $\nu_e$ due to its oscillating into $\nu_s$ after decoupling from the background plasma increases the $^4$He yield, whereas the corresponding depletion of $\nu_\mu$ has no effect on $^4$He production, since $\nu_\mu$ does not interact directly with the protons and neutrons.

The bounds from LSS are also shifted by the presence of the primordial asymmetry. For a given $\Delta m^2_{14}$ and a given $\theta_s$, the lower final abundances of sterile neutrinos brought about by the non-zero asymmetry allows for a larger sterile neutrino mass for a fixed $\Omega_\nu h^2$, resulting in a less stringent constraint.

\begin{figure}[t]
$$
\includegraphics[width=0.60\textwidth]{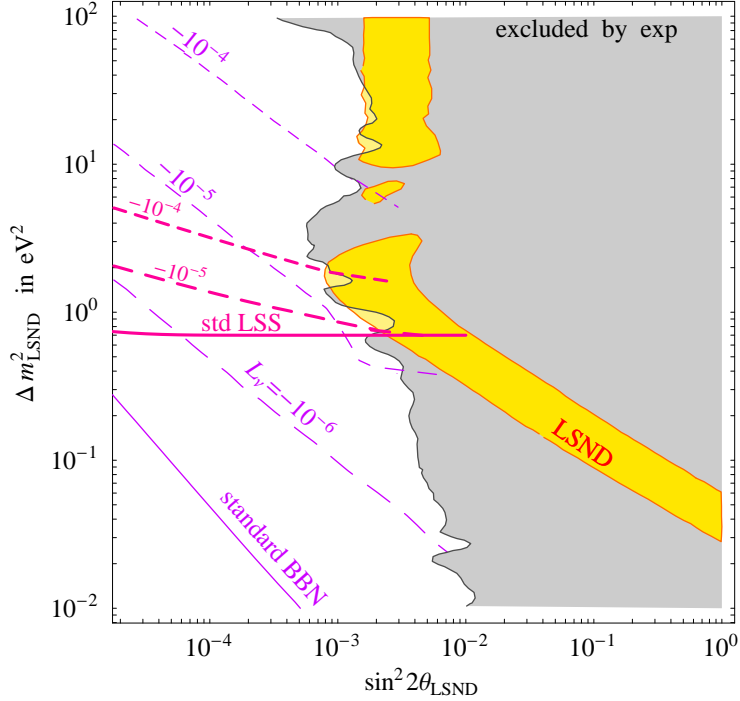}$$
\caption[X]{\label{fig:LSND}\small The allowed LSND region at 99\% C.L. (yellow/light shaded area) compared to the cosmological bounds from BBN and LSS in the presence of primordial asymmetries. The darker shaded area is already excluded at 99\% C.L. by other experiments. The regions below and to the left of the thin lines are allowed by BBN because they correspond to $Y_p \le 0.258$. The regions below and to the left of the thick lines are allowed by LSS because they correspond to $\Omega_\nu h^2 \le 0.8\ 10^{-2}$.}
\end{figure}

\subsection{Rescuing LSND}
\label{sec:LSND}

The LSND experiment reported a signal \cite{LSND} for $\bar\nu_\mu \to \bar\nu_e$ oscillations in the appearance of $\bar\nu_e$ in an originally $\bar\nu_\mu$ beam. The best fit point is located at $\sin^2\theta_{\rm LSND} = 3 \cdot \, \, 10^{-3}$, $\Delta m^2_{\rm LSND} = 1.2 \eV^2$ (the whole allowed region is represented in fig.\ref{fig:LSND} as a lightly colored area).  
The ``3+1 sterile'' neutrino explanation assumes that the $\bar\nu_\mu \to \bar\nu_e$ oscillation proceeds through $\bar\nu_\mu \to \bar\nu_s \to \bar\nu_e$, therefore introducing the large LSND mass scale $\Delta m^2_{\rm LSND}$ as the mass squared difference of the active neutrinos with the mainly sterile state. The effective LSND mixing angle can be expressed in terms of the active/sterile mixing angles $\theta_{e{\rm s}}$ and $\theta_{\mu{\rm s}}$ as~\cite{LSND others}
\beq
\label{LSND angles}
\sin^2 2\theta_{\rm LSND} = \frac{1}{4} \sin^2 2\theta_{e{\rm s}} \sin^2 2\theta_{\mu{\rm s}},
\eeq
which reduces to $\theta_{\rm LSND} \approx\theta_{e{\rm s}}\cdot \theta_{\mu{\rm s}}$ for small mixing.
Both angles are constrained by several other experiments that found no evidence of electron or muon neutrino disappearance. As a result~\cite{LSND strumia}, a large portion of the area indicated by LSND, but not all of it, is ruled out (the darker shaded area in fig.~\ref{fig:LSND}).

The constraint by cosmology are represented by lines in fig.~\ref{fig:LSND}. They have been computed in the most conservative way, i.e. by choosing for each value of $\sin^22\theta_{\rm LSND}$ the combination $\theta_{e{\rm s}}\approx\theta_{\mu{\rm s}}\approx\theta_{\rm LSND}^{1/2}$, which is found to minimize the overproduction of sterile neutrinos. When these values are ruled out by neutrino data, we set one of the two angles to its largest allowed value and determine the other using eq. (\ref{LSND angles}). On physical grounds, setting $\theta_{e{\rm s}}\approx\theta_{\mu{\rm s}}$ means the $es$ and $\mu s$ "channels" both contribute equally to the production of $^4$He and Deuterium. From eq. (\ref{LSND angles}), this implies that the LSND plot in fig. (\ref{fig:LSND}) can be obtained, in a qualitative sense, from fig. (\ref{fig:esmus}) by a re-scaling $\sin^22 \theta \to 1/4\ (\sin^2 2 \theta)^2$ (for small mixing). This guideline is expected to hold for other mixing patterns that involve two flavors with similar mixing angles. 

We plot the isocontours of $Y_p = 0.258$ and $\Omega_\nu h^2 = 0.8\ 10^{-2}$, so that the regions above and to the right of these lines are the excluded portions. One recovers the well known result that the region excluded by standard BBN ($L_\nu = 0$, solid thin line) includes all of the allowed LSND region. The bound from LSS (solid thick line), updated by the recent cosmological results as discussed in Sec.\ref{sec:LSS}, now also lies below the LSND allowed portions.

The effect of introducing the primordial asymmetry is represented by the dashed lines: the larger the value of the (negative) asymmetry, the more inhibited the production of sterile neutrinos, and the further upper right the $^4$He and $\Omega_\nu h^2$ contours are displaced.
We find that a primordial asymmetry $L_\nu$ of the order of $-10^{-4}$ is enough to re-allow the portion of the LSND region around $(\Delta m^2_{\rm LSND},\ \sin^22\theta_{\rm LSND}) \simeq (1\ {\rm eV}^2, {\rm few}\ 10^{-3})$. This result is less stringent than the estimate of $L_{\nu_e} \simeq {\cal O}(10^{-1 \ldots 2})$ in~\cite{BargerHiding}, that was derived without the use of the explicit neutrino evolution equations.

\section{Conclusions}
\label{conclusions}

We investigated the extent to which introducing a primordial leptonic asymmetry $(n_{\nu} \neq n_{\bar\nu})$, much larger than the baryonic one, relaxes the bounds on light sterile neutrinos imposed by Big Bang Nucleosynthesis and Large Scale Structure, by suppressing via matter effects the production of the sterile states in the Early Universe. 
Our momentum-averaged neutrino evolution equations took into account the background cosmological evolution and included all the relevant features of neutrino mixing, both active-active and active-sterile, in the presence of a primordial leptonic asymmetry.

Working in the framework of constant leptonic asymmetry, we solved numerically the neutrino densities as a function of temperature and computed the relevant cosmological observables such as the yield of $^4$He and Deuterium and the neutrino contribution to the energy density in the Universe $\Omega_\nu h^2$, comparing them with the most recent determinations.

We applied this formalism to the study of the generic cases of a sterile neutrino mixing with a pure electron neutrino and with a pure muon one. For different values of the primordial asymmetry $L_\nu$, taken to be the same for all the active flavors, we found the relaxed constraints on the parameter spaces. The results are presented in Fig.~\ref{fig:esmus} across a broad range of mixing parameters.

We then studied the specific case of the LSND sterile neutrino, which mixes with a specific combination of electron and muon neutrinos. The constraint from standard cosmology in the absence of asymmetries are confirmed to rule out the entire LSND region, by both BBN $^4$He overproduction and the cosmological neutrino mass bound. We found that a primordial asymmetry of order $\mathcal{O}(10^{-4})$ allows the LSND regions around $(\Delta m^2_{\rm LSND}, \sin^22\theta_{\rm LSND}) \simeq (1\ {\rm eV}^2, {\rm few}\ 10^{-3})$ to be reopened.

\appendix

\section{Appendix}
\label{appendix}
In this Appendix we lay down explicitly the formalism that we use for the evolution of the temperature in our computation. In full generality, a standard analysis~\cite{enqvist,weinberg} gives for the differential change of temperature
\beq
\label{eq:Tchange}
\frac{dT}{dt} = \left( \frac{\partial T}{\partial a} \right)_{N_\nu} \dot{a} + \left( \frac{\partial T}{\partial N_\nu} \right)_{a}  \dot{N_\nu}
\eeq
where the first term accounts for the expansion of the Universe ($a$ being the scale factor) and the second term for the variation in the count of the neutrino total degrees of freedom $N_\nu$. The first term can be computed exploiting the conservation of the entropy density in the system of photons and $e^\pm$ per comoving volume 
\begin{align}
\globallabel{eq:entropy}
s =& \frac{a^3}{T} \left( \rho_\gamma + \rho_{e^\pm} + P_\gamma + P_{e^\pm} \right) = (aT)^3 \frac{4 \pi^2}{45} \mathcal{S}\left(\frac{m_e}{T}\right), \mytag \\
 &{\rm with}\ \  \mathcal{S}(x) = 1 + \frac{45}{2\pi^4} \int {\rm d}y\ y^2 \left( \sqrt{y^2+x^2} + \frac{y^2}{3 \sqrt{y^2+x^2} }\right)  \frac{1}{e^{\sqrt{y^2+x^2}}+1} \mytag
\end{align}
that effectively provides an $a \leftrightarrow T$ relation. One obtains (here $\mathcal{S} \equiv \mathcal{S}(m_e/T)$)
\beq
\label{eq:dTda}
\left( \frac{\partial T}{\partial a} \right)_{N_\nu}  \dot{a} = \frac{\dot{a}}{a} 
\left( \frac{-T\mathcal{S}^{1/3}}{\frac{\partial}{\partial T}(T\mathcal{S}^{1/3})} \right) = H(\wp_{\rm tot}) \left( \frac{-T\mathcal{S}^{1/3}}{\frac{\partial}{\partial T}(T\mathcal{S}^{1/3})} \right)
\eeq
In the same way, one obtains the relation between $T_\nu$ and $T$
\beq
\left( \frac{T_\nu}{T}\right)^3 = \frac{4}{11}\mathcal{S}.
\eeq
$\mathcal{S}(m_e/T)-1$ effectively parameterizes the entropy in electrons and positrons at any given $T$, including during their annihilation into photons; for $T \gg m_e$, $\mathcal{S} \to 11/4$.

The second term in eq.\eq{Tchange} can be computed exploiting the conservation of the total energy density (at fixed $a$)
\begin{align}
\globallabel{eq:totalenergydensity}
\wp_{\rm tot} =& \left( \rho_\gamma + \sum \rho_{\nu,\nub} + \rho_{e^\pm}  \right) = \frac{\pi^2}{30} T^4 \left( 2+\frac{7}{8}N_\nu \left( \frac{T_\nu}{T} \right)^4  +\mathcal{R}\left(\frac{m_e}{T}\right) \right), \mytag \\
 &{\rm with}\ \  \mathcal{R}(x) = \frac{60}{\pi^4} \int {\rm d}y\ y^2 \sqrt{y^2+x^2}   \frac{1}{e^{\sqrt{y^2+x^2}}+1} \mytag
\end{align}
that effectively provides an $N_\nu \leftrightarrow T$ relation. One obtains the lengthy expression (here $\mathcal{R} \equiv \mathcal{R}(m_e/T)$)
\beq
\left( \frac{\partial T}{\partial N_\nu} \right)_{a}  \dot{N_\nu} = - \frac{\dot{N_\nu} T}{4} 
\left( N_\nu \left( 1+ \frac{T}{3 \mathcal{S}} \frac{\partial \mathcal{S}}{\partial T} \right)
+ \frac{8}{7} \left(\frac{11}{4} \right)^{4/3} \mathcal{S}^{-4/3} \left( 1+ \frac{\mathcal{R}}{2} +\frac{T}{8} \frac{\partial \mathcal{R}}{\partial T} \right) \right)^{-1}
\eeq
In our case, this contribution, proportional to $\dot{N_\nu}$, is subdominant with respect to the first one in eq.\eq{Tchange}, since the differential change in $N_\nu$ due to the production of the fourth family of neutrinos is limited by their contribution to the total energy budget of the universe, which is at most $\mathcal{O}(7/50)$. This effect is also already included in the expression for the Hubble parameter $H(\wp_{\rm tot})$.
Therefore in the $n/p$ evolution equation eq.\eq{noverp} we only keep the first addend of the $dT/dt$ relation eq.\eq{Tchange}. 
Finally, we note that in the neutrino evolution equations eq.s\eq{kineq nu} we have made use of the further approximated expression $dT/dt =-H(\wp_{\rm tot})T$. Eq.s \eq{Tchange} and \eq{dTda} reduce to this form at $T \gg m_e$ when $\mathcal{S}$ reduces to the constant value $11/4$. In this same regime $\mathcal{R} \to 7/2$ and of course $T_\nu = T$.

\footnotesize

\paragraph{Acknowledgments.}
We are especially grateful to Guido Marandella for useful discussions and support, and to Alessandro Curioni, Alessandro Strumia and Francesco Vissani. The work of M.C. was supported in part by the USA DOE-HEP Grant DE-FG02-92ER-40704; up to May 2006, the work of Y.-Z. C. was supported by the same grant.


\begin{thebibliography}{99}

\bibitem{historical}
Sterile/active oscillations and BBN:
D. Kirilova, Dubna preprint JINR E2-88-301.
  R.~Barbieri and A.~Dolgov,
  Phys.\ Lett.\ B {\bf 237}, 440 (1990).
K.~Enqvist, K.~Kainulainen and J.~Maalampi,
  Phys.\ Lett.\ B {\bf 249} (1990) 531.
  K.~Kainulainen,
  Phys.\ Lett.\ B {\bf 244}, 191 (1990).
  R.~Barbieri and A.~Dolgov,
  Nucl.\ Phys.\ B {\bf 349}, 743 (1991).
  K.~Enqvist, K.~Kainulainen and M.~J.~Thomson,
  Nucl.\ Phys.\ B {\bf 373}, 498 (1992).
    J.~M.~Cline,
  Phys.\ Rev.\ Lett.\  {\bf 68}, 3137 (1992).
X.~Shi, D.~N.~Schramm and B.~D.~Fields,
  Phys.\ Rev.\ D {\bf 48}, 2563 (1993)
  [arXiv:astro-ph/9307027].
  E.~Lisi, S.~Sarkar and F.~L.~Villante,
  Phys.\ Rev.\ D {\bf 59}, 123520 (1999)
  [arXiv:hep-ph/9901404].
K.~N.~Abazajian,
  Astropart.\ Phys.\  {\bf 19}, 303 (2003)
  [arXiv:astro-ph/0205238].

  
\bibitem{LesgourguesPastorReview}
See e.g. 
  J.~Lesgourgues and S.~Pastor,
  Phys.\ Rept.\  {\bf 429} (2006) 307
  [arXiv:astro-ph/0603494].


\bibitem{LSND} A.~Aguilar {\it et al.}  [LSND Collaboration],
Phys.\ Rev.\ D {\bf 64} (2001) 112007
[arXiv:hep-ex/0104049].


\bibitem{CMSV} 
M.~Cirelli, G.~Marandella, A.~Strumia and F.~Vissani,
  Nucl.\ Phys.\ B {\bf 708} (2005) 215
  [arXiv:hep-ph/0403158].
See also references therein for additional literature. 
Aspects of the study were discussed in 
M.~Cirelli,
  arXiv:astro-ph/0410122,
G.~Marandella,
  arXiv:hep-ph/0405090
and
A.~Strumia,
  Nucl.\ Phys.\ Proc.\ Suppl.\  {\bf 143}, 144 (2005)
  [arXiv:hep-ph/0407132].


\bibitem{cosmonu}
M.~Cirelli and A.~Strumia,
  arXiv:astro-ph/0607086.


\bibitem{di bari}
P.~Di Bari,
  Phys.\ Rev.\ D {\bf 65} (2002) 043509
  [Addendum-ibid.\ D {\bf 67} (2003) 127301]
  [arXiv:hep-ph/0108182].
  
  
  \bibitem{LSS}
A.~Strumia,
  Phys.\ Lett.\ B {\bf 539} (2002) 91
  [arXiv:hep-ph/0201134].
    A.~Pierce and H.~Murayama,
  Phys.\ Lett.\ B {\bf 581}, 218 (2004)
  [arXiv:hep-ph/0302131].
C.~Giunti,
  Mod.\ Phys.\ Lett.\ A {\bf 18}, 1179 (2003)
  [arXiv:hep-ph/0302173].
  S.~Hannestad,
  JCAP {\bf 0305}, 004 (2003)
  [arXiv:astro-ph/0303076].
  More recently, the implication of LSS measurement on sterile neutrino masses are addressed in
  S.~Dodelson, A.~Melchiorri and A.~Slosar,
  Phys.\ Rev.\ Lett.\  {\bf 97}, 04301 (2006)
  [arXiv:astro-ph/0511500]
  and in~\cite{cosmonu}.
  
\bibitem{miniboone}
See the ``MiniBooNE Run Plan'' (November 2003) available from http://www-boone.fnal.gov/publicpages.

\bibitem{rprocess}
J.~Beun, G.~C.~McLaughlin, R.~Surman and W.~R.~Hix,
  Phys.\ Rev.\ D {\bf 73}, 093007 (2006)
  [arXiv:hep-ph/0602012].

\bibitem{bounds on xi}
H.~S.~Kang and G.~Steigman,
  Nucl.\ Phys.\ B {\bf 372}, 494 (1992).
C.~Lunardini and A.~Y.~Smirnov,
Phys.\ Rev.\ D {\bf 64} (2001) 073006
[arXiv:hep-ph/0012056].
A.~D.~Dolgov, S.~H.~Hansen, S.~Pastor, S.~T.~Petcov, G.~G.~Raffelt and D.~V.~Semikoz,
Nucl.\ Phys.\ B {\bf 632} (2002) 363
[arXiv:hep-ph/0201287].
K.~N.~Abazajian, J.~F.~Beacom and N.~F.~Bell,
  Phys.\ Rev.\ D {\bf 66}, 013008 (2002)
  [arXiv:astro-ph/0203442].
  Y.~Y.~Y.~Wong,
  Phys.\ Rev.\ D {\bf 66}, 025015 (2002)
  [arXiv:hep-ph/0203180].
  More recently, these bounds have been updated in:  
  A.~Cuoco, F.~Iocco, G.~Mangano, G.~Miele, O.~Pisanti and P~.D.~Serpico,
  Int. J. Mod. Phys. A 19, 4431 (2004)
  [arXiv:astro-ph/0307213].
  P.~D.~Serpico and G.~G.~Raffelt,
  Phys. Rev. D  71, 127301 (2005)
  [arXiv:astro-ph/0506162].
  
  
  \bibitem{scenariosLargeL}
  Large primordial leptonic asymmetries in cosmological scenarios:
  J.~A.~Harvey and E.~W.~Kolb,
  Phys.\ Rev.\ D {\bf 24}, 2090 (1981).
  A.~Casas, W.~Y.~Cheng and G.~Gelmini,
  Nucl.\ Phys.\ B {\bf 538} (1999) 297
  [arXiv:hep-ph/9709289].
  J.~March-Russell, H.~Murayama and A.~Riotto,
  JHEP {\bf 9911}, 015 (1999)
  [arXiv:hep-ph/9908396].
   J.~McDonald,
  Phys.\ Rev.\ Lett.\  {\bf 84}, 4798 (2000)
  [arXiv:hep-ph/9908300].
  M.~Kawasaki, F.~Takahashi and M.~Yamaguchi,
  Phys.\ Rev.\ D {\bf 66}, 043516 (2002)
  [arXiv:hep-ph/0205101].

\bibitem{original Foot Volkas} 
R.~Foot and R.~R.~Volkas,
  Phys.\ Rev.\ Lett.\  {\bf 75}, 4350 (1995)
  [arXiv:hep-ph/9508275].

  \bibitem{asymm}
Sterile/active oscillations and neutrino asymmetry:
R.~Foot, M.~J.~Thomson and R.~R.~Volkas,
  Phys.\ Rev.\ D {\bf 53}, 5349 (1996)
  [arXiv:hep-ph/9509327].
  D.~P.~Kirilova and M.~V.~Chizhov,
  Phys.\ Lett.\ B {\bf 393} (1997) 375
  [arXiv:hep-ph/9608270].
D.~P.~Kirilova and M.~V.~Chizhov,
  Phys.\ Rev.\ D {\bf 58}, 073004 (1998)
  [arXiv:hep-ph/9707282].
D.~P.~Kirilova and M.~V.~Chizhov,
  Nucl.\ Phys.\ B {\bf 534} (1998) 447
  [arXiv:hep-ph/9806441].
  D.~P.~Kirilova and M.~V.~Chizhov,
  Nucl.\ Phys.\ B {\bf 591}, 457 (2000)
  [arXiv:hep-ph/9909408].

\bibitem{BargerHiding} V.~Barger, J.~P.~Kneller, P.~Langacker, D.~Marfatia and G.~Steigman,
Phys.\ Lett.\ B {\bf 569} (2003) 123
[arXiv:hep-ph/0306061].

\bibitem{dolgov villante}
A.~D.~Dolgov and F.~L.~Villante,
  Nucl.\ Phys.\ B {\bf 679} (2004) 261
  [arXiv:hep-ph/0308083].


\bibitem{fuller} K.~Abazajian, N.~F.~Bell, G.~M.~Fuller and Y.~Y.~Y.~Wong,
Phys.\ Rev.\ D {\bf 72} (2005) 063004
[arXiv:astro-ph/0410175].
C.~T.~Kishimoto, G.~M.~Fuller and C.~J.~Smith,
  arXiv:astro-ph/0607403.


\bibitem{StrumiaVissaniReview}
A.~Strumia and F.~Vissani,
  arXiv:hep-ph/0606054.


\bibitem{WMAP3}
D.~N.~Spergel {\it et al.},
  arXiv:astro-ph/0603449.


\bibitem{OscEU}
Neutrino oscillations in the Early Universe:
A.~D.~Dolgov,
  Sov.\ J.\ Nucl.\ Phys.\  {\bf 33}, 700 (1981)
  [Yad.\ Fiz.\  {\bf 33}, 1309 (1981)].
L.~Stodolsky,
  Phys.\ Rev.\ D {\bf 36}, 2273 (1987).
A.~Manohar,
  Phys.\ Lett.\  {\bf 186B}, 370 (1987).
M.~J.~Thomson and B.~H.~J.~McKellar,
  Phys.\ Lett.\ B {\bf 259}, 113 (1991).
J.~T.~Pantaleone,
  Phys.\ Lett.\ B {\bf 287}, 128 (1992).
A.~Friedland and C.~Lunardini,
  Phys.\ Rev.\ D {\bf 68}, 013007 (2003)
  [arXiv:hep-ph/0304055].
  
  
\bibitem{notzold}
D.~Notzold and G.~Raffelt,
  Nucl.\ Phys.\ B {\bf 307} (1988) 924.

\bibitem{sigl raffelt}
G.~Sigl and G.~Raffelt,
  Nucl.\ Phys.\ B {\bf 406}, 423 (1993).

\bibitem{dolgov review}
A.~D.~Dolgov,
  Phys.\ Rept.\  {\bf 370} (2002) 333
  [arXiv:hep-ph/0202122].


\bibitem{xi in CMB} M.~Lattanzi, R.~Ruffini and G.~V.~Vereshchagin,
Phys.\ Rev.\ D {\bf 72} (2005) 063003
[arXiv:astro-ph/0509079].
 (Table 1) and 
 J.~Lesgourgues and S.~Pastor,
Phys.\ Rev.\ D {\bf 60}, 103521 (1999)
[arXiv:hep-ph/9904411].

\bibitem{enqvist}
K.~ Enqvist as cited in \cite{historical}.

\bibitem{kirilova2005}
D.~P.~Kirilova,
  arXiv:astro-ph/0511231.
  D.~P.~Kirilova and M.~P.~Panayotova,
  arXiv:astro-ph/0608103.

\bibitem{kirilova} D.~Kirilova et al. in \cite{asymm}.

\bibitem{rising of asy}
R.~Foot {\it et al.} in~\cite{asymm}.
R.~Foot, M.~J.~Thomson and R.~R.~Volkas,
  Phys.\ Rev.\ D {\bf 53} (1996) 5349
  [arXiv:hep-ph/9509327].
  R.~Foot and R.~R.~Volkas,
  Phys.\ Rev.\ D {\bf 55} (1997) 5147
  [arXiv:hep-ph/9610229].
  R.~Foot and R.~R.~Volkas,
  Phys.\ Rev.\ D {\bf 56} (1997) 6653
  [Erratum-ibid.\ D {\bf 59} (1999) 029901]
  [arXiv:hep-ph/9706242].
  N.~F.~Bell, R.~Foot and R.~R.~Volkas,
  Phys.\ Rev.\ D {\bf 58} (1998) 105010
  [arXiv:hep-ph/9805259].
  R.~Foot,
  Astropart.\ Phys.\  {\bf 10} (1999) 253
  [arXiv:hep-ph/9809315].
  R.~Foot,
  Phys.\ Rev.\ D {\bf 61} (2000) 023516
  [arXiv:hep-ph/9906311].
   P.~Di Bari, P.~Lipari and M.~Lusignoli,
  Int.\ J.\ Mod.\ Phys.\ A {\bf 15} (2000) 2289
  [arXiv:hep-ph/9907548].
  A.~D.~Dolgov, S.~H.~Hansen, S.~Pastor and D.~V.~Semikoz,
  Astropart.\ Phys.\  {\bf 14} (2000) 79
  [arXiv:hep-ph/9910444].
  P.~Di Bari, R.~Foot, R.~R.~Volkas and Y.~Y.~Y.~Wong,
  Astropart.\ Phys.\  {\bf 15} (2001) 391
  [arXiv:hep-ph/0008245].
  A.~D.~Dolgov,
  Nucl.\ Phys.\ B {\bf 610} (2001) 411
  [arXiv:hep-ph/0102125].
  P.~Di Bari and R.~Foot,
  Phys.\ Rev.\ D {\bf 65} (2002) 045003
  [arXiv:hep-ph/0103192].


\bibitem{esposito}
S.~Esposito, G.~Mangano, G.~Miele and O.~Pisanti,
  Nucl.\ Phys.\ B {\bf 540}, 3 (1999)
  [arXiv:astro-ph/9808196].

\bibitem{PDG} 
S.~Eidelman {\it et al.}  [Particle Data Group],
  Phys.\ Lett.\ B {\bf 592}, 1 (2004).

\bibitem{PDGSarkar}
B.~Fields and S.~Sarkar,
  arXiv:astro-ph/0601514.

\bibitem{OliveSkillman}
 K.~A.~Olive and E.~D.~Skillman,
  Astrophys.\ J.\  {\bf 617} (2004) 29
  [arXiv:astro-ph/0405588].
  
\bibitem{Kirkman}
D.~Kirkman, D.~Tytler, N.~Suzuki, J.~M.~O'Meara and D.~Lubin,
  Astrophys.\ J.\ Suppl.\  {\bf 149} (2003) 1
  [arXiv:astro-ph/0302006].

\bibitem{proceeding}
M.~Cirelli as cited in~\cite{CMSV}. Recently, 
M.~Fukugita and M.~Kawasaki,
  arXiv:astro-ph/0603334
has also obtained a value in line with~\cite{OliveSkillman}.

\bibitem{omeganu}
S.~Dodelson {\it et al.} as cited in~\cite{LSS}, and Ref.~\cite{cosmonu}.

\bibitem{summnu}
A.~Goobar, S.~Hannestad, E.~Mortsell and H.~Tu,
  arXiv:astro-ph/0602155.
D.~N.~Spergel in~\cite{WMAP3}.  
M.~Fukugita and M.~Kawasaki,
  arXiv:astro-ph/0603334.
  U.~Seljak, A.~Slosar and P.~McDonald,
  arXiv:astro-ph/0604335.
G.~L.~Fogli {\it et al.},
  arXiv:hep-ph/0608060.
  
  \bibitem{BD}
  R.~Barbieri and A.~Dolgov in~\cite{historical}.

\bibitem{LSND others}
N.~Okada and O.~Yasuda,
  Int.\ J.\ Mod.\ Phys.\ A {\bf 12} (1997) 3669
  [arXiv:hep-ph/9606411].
S.~M.~Bilenky, C.~Giunti and W.~Grimus,
  Eur.\ Phys.\ J.\ C {\bf 1} (1998) 247
  [arXiv:hep-ph/9607372].

\bibitem{LSND strumia}
A.~Strumia as cited in~\cite{LSS}.

\bibitem{weinberg}
S.~Weinberg, ``Gravitation and Cosmology: Principles and Applications of the General Theory of Relativity'', J.Wiley and Sons, July 1972; Chapter 15.6.


\end{thebibliography}
\end{document}